\def\beg{\begin{equation}}
\def\eeq{\end{equation}}
\documentstyle[12pt]{article}
\begin{document}
\begin{center}
{\Large{\bf Clustering in quantum Hall effect: Spin-charge coupling}}
\vskip0.55cm
{\bf Keshav N. Shrivastava}
\vskip0.25cm
{\it School of Physics, University of Hyderabad,\\
Hyderabad  500046, India}
\end{center}
     The effective fractional charges like 17/4 or 19/4 are explained by our angular momentum theory. These fractions do not arise from odd-denominator rule. Due to same spin polarization for both of these along the magnetic field, these states are not the particle-hole conjugates. The idea of clustering first introduced in cond-mat/0303309 has been extended to atomic clusters which explain the oscillations in the kHz range.
\vskip1.0cm
Corresponding author: keshav@mailaps.org\\
Fax: +91-40-2301 0145.Phone: 2301 0811.
\vskip1.0cm

\noindent {\bf 1.~ Introduction}

     Recently, Cooper et al have obtained signals of frequency varying from about 3 kHz to 25 kHz when the d.c. voltage is increased such that the d.c. current increases beyond a threshold value. The conditions for the plateaus in the quantum Hall effect are obtained at low d.c. voltage. When the d.c. voltage is increased  the plateau disappears and a finite resistivity is obtained. The phase change from zero resistivity to finite resistivity by increasing voltage is called the reentrant quantum Hall effect. In the finite resistivity phase there is a kHz signal. Such kHz signals are not obtained in all of the plateaus but are characteristic of only some of the plateaus. In the case of two of the
plateaus, where experimental measurements of kHz oscillations have been made, the particle-hole symmetry could not be identified and these two states are not particle-hole conjugates. When some effort is made, the predicted frequencies occur in the range of MHz or GHz whereas the experimental values are of the order of kHz. Therefore, it may be assumed that the origin of the kHz oscillations has not been understood.

     In the present letter, we report that the kHz oscillations are caused by ``clustering". We can predict their frequency and their symmetry correctly.

\noindent{\bf 2.~~Theory}

     We consider that the sample of the size of a few mm consists of clusters of atoms. Every cluster has a finite number of polarized electrons. Some of the clusters have only one electron and only one atom so that these clusters do not contribute to kHz  oscillations but they do exhibit particle-hole symmetry. Therefore, we consider two types of clusters, one with only one atom, and the other with a finite number of atoms. These clusters are inferred from the spin value. For spin=1/2, we have clusters with only one atom and for $s>1/2$ we have clusters with a large number of atoms per cluster. First we consider clusters with only one atom per cluster or with spin 1/2 only.

     The quantum Hall effect has always been a subject with many different theories. Our theory makes use of the theory of angular momentum[2-9]. Recently, Pan et al[10] have reported observation of some of the fractions which could not be obtained by their model. However, our theory explains  these values very well, particularly when $s>1/2$ has been considered[11]. It was found that Mani and von Klitzing[12] have made a very detailed measurement of fractions which determine the effective charge. We are able to predict all of the 146 measured effective fractional values[13]. Similarly, Smet[14] also reported that the fractions were not the same as those found by their model. These fractions are also predicted by our theory correctly [15]. A study of cyclotron resonance which gave new splitting could not be understood by their model[16] but our theory explains the data very well[17]. The thermoelectric power in between two plateaus is also well predicted by our theory[18].

     The fractional charge derived by us by using the theory of angular momentum is given by,
\beg
e_{eff}/e= {{\it l}+{1\over 2}\pm s\over 2{\it l}+1}
\eeq
for $s=\pm 1/2$, for various values of ${\it l}$ these predicted values are the same as those measured [19]. The details of derivation are similar to that of Lande's g-value formula except that Lande used only the + sign in $j={\it l}\pm s$, whereas we use both the plus as well as the minus sign. When ${\it l}$=0, $s$=1/2 we get $e_{eff}/e$=1 and for ${\it l}$=1, $s$=-1/2, we get 1/3. As long as $s$=1/2, there is no clustering or we can say that there is only one electron in a cluster. This also means that there is only one atom in a cluster. For large values of $s$, there is a large number of electrons which are polarized in a magnetic field. Similarly, a large number of electrons also means that there are a lot of atoms in a cluster.

     We analyse the two fractions and find their symmetry. The fraction 4 + (1/4)=17/4 may be analysed as $(n+{1\over 2})$17/2 where $n$ is the Landau level quantum number. We take $n$=0 so that at low temperatures, the fraction is (17/2)(1/2). For ${\it l}$=0 in the above formula, the effective charge is,
\beg
{1\over 2}\pm s = {17\over 2}
\eeq
for $s$=8 and positive sign, we obtain 8+(1/2)=17/2. We consider only the positive sign for only one sign of the fractional charge. The spin, s=8 means 16/2, so that there are atleast 16 electrons in the cluster. Pauli principle will make sure that they are not in one atom. This shows that there are electron clusters. Similarly, 4+(3/4) =19/4 can be written as (1/2)(19/2) so that the effective charge is,
\beg
{1\over 2}\pm s ={19\over 2}
\eeq
so that for positive sign $s$=9 which is 18/2 so that there is a cluster of atleast 18 electrons. This shows that there is a cluster. Since for 17/2 as well as for 19/2 we used positive sign, there are no values with negative sign. The negative sign values are the partile-hole symmetric states which are conjugate to those with positive sign. Therefore 19/4 as well as 17/4 are both spin polarized in one and the same direction and there are no conjugate states. Therefore 17/4 is not the conjugate of 19/4.

     The $n$=0 in $(n+{1\over 2})\hbar\omega_c$ produces only the zero-point frequency with a factor of 1/2 so that 25/4 reduces to 25/2. In the formula (1) we substitute ${\it l}$=0 so that,
\beg
{1\over 2}\pm s ={25\over 2}
\eeq
or s=12 with positive sign. That is possible only for 24 electrons which form a cluster. Using the negative sign, we can obtain the Kramers conjugate states. For 25/2 the conjugate is -23/2; for 17/2 it is -15/2 and for 19/2 it is -17/2. The conjugate states require the spin-reversed states which are not populated due to strong magnetic field.

     For spin 3/2, there are atleast three electrons.  Since {\it l}=0, three sites are needed. For spin 5/2, five sites are needed but a spin-charge relationship is clearly predicted,
\beg
{1\over 2}= charge \mp spin.
\eeq

     At very low temperatures, the velocity of an atom measures the thermodynamic temperature,
\beg
{1\over 2}mv^2=3 k_B T
\eeq
one each $k_BT$ for x, y and z components of the velocity. If the temperature is T=50 mK, $k_B=1.38\times 10^{-16}\,\,\,erg/K$, the mass of the cluster, $m =N\times 69.723\times 1.67261\times 10^{-24} \,\,g$ with N the number of atoms $N\simeq 4\times 10^7$ for the quantum well of size 30 nm, we obtain,$ v=0.094 \,\,cm/s$ and for $d=30 \,\,nm$,
\beg
{v\over d} = 31 \times 10^3 Hz
\eeq
which is of the order of kHz. This predicted frequency is of the correct order of magnitude considering the uncertainty in the number of atoms in the cluster. If the resistivity is fluctuating, it is possible that the system may radiate in the kHz region.

\noindent{\bf3.~~ Conjugate states}.

     The fractions such as 1/3 and 2/3 belong to $s=\pm 1/2$. They do not belong to clusters. The above phenomenon belongs to clusters. Therefore, 1/3, 2/3 do not produce oscillations in the kHz range. However, 17/2 and 19/2 belong to clusters and hence give kHz oscillations. When voltage is varied 19/2 switches from plateau to finite resistivity state but 1/3 does not. That is why 19/2 gives a reentrant quantum Hall effect (RQHE). The states 17/2 and 19/2 both use only the positive sign for spin. Hence these are not the Kramers conjugate states.

\noindent{\bf4.~~ Conclusions.}

     The fractions 17/4 and 19/4 represent spin polarized states. These are not the particle-hole conjugates. The fractions like 1/3 represent only one particle whereas 17/4 comes from a cluster. Therefore, 1/3 does not produce noise but 17/4 does. We find that spin and charge are linearly related.

\noindent{\bf5.~~References}
\begin{enumerate}
\item K. B. Cooper, J. P. Eisenstein, L. N. Pfeiffer and K. W. West, Phys. Rev. Lett. {\bf 90}, 226803 (2003).
\item K. N. Shrivastava, Phys. Lett. A {\bf 113}, 435 (1986).
\item K. N. Shrivastava, Mod. Phys. Lett. B {\bf 13}, 1087 (1999).
\item K. N. Shrivastava, Mod. Phys. Lett. B {\bf 14}, 1009 (2000).
\item K. N. Shrivastava, in Frontiers of Fundamental Physics 4, B. G. Sidharth and M. V. Altaisky, Kluwer Academic/Plenum Pub. New York 2001.
\item K. N. Shrivastava, Superconductivity: Elementary Topics, World Scientific, New Jersey, London, Singapore, 2000.
\item K.N. Shrivastava, Introduction to quantum Hall effect,\\ 
      Nova Science Pub. Inc., N. Y. (2002).
\item K. N. Shrivastava, Natl. Acad. Sci. Lett. (India) {\bf 9}, 145 (1986).
\item K. N. Shrivastava, Natl. Acad. Sci. Lett. (India) (2003).
\item W. Pan, H. L. Stormer, D. C. Tsui, L. N. Pfeiffer, K.  W. Baldwin and K. W. West, Phys. Rev. Lett. {\bf 90}, 016801 (2003).
\item K. N. Shrivastava, cond-mat/0302610.
\item R. Mani and K. von Klitzing, Z. Phys. B {\bf 100}, 635 (1996).
\item K. N. Shrivastava, cond-mat/0303309 and cond-mat/0303621.
\item J. H. Smet, Nature, {\bf 422}, 391 (2003).
\item K. N. Shrivastava, cond-mat/0304269.
\item S. Syed, M. J. Manfra, Y. J. Wang, H. L. Stormer and R. J. Molner, cond-mat/0305358.
\item K. N. Shrivastava, cond-mat/0305415.
\item K. N. Shrivastava, cond-mat/0305379.
\item H. L. Stormer, Rev. Mod. Phys. {\bf 71}, 875 (1999).
\end{enumerate}
\vskip0.1cm

Note: Ref.7 is available from:\\
 Nova Science Publishers, Inc.,\\
400 Oser Avenue, Suite 1600,\\
 Hauppauge, N. Y.. 11788-3619,\\
Tel.(631)-231-7269, Fax: (631)-231-8175,\\
 ISBN 1-59033-419-1 US$\$69$.\\
E-mail: novascience@Earthlink.net

\end{document}